\documentclass[aps,prl,reprint,superscriptaddress,floatfix,flushbottom,nobibnotes,nofootinbib,a4paper,showpacs,amsmath]{revtex4-1}
\usepackage{amssymb}
\usepackage{graphicx}
\newcommand{\be}{\begin{equation}}
\newcommand{\ee}{\end{equation}}
\newcommand{\bea}{\begin{eqnarray}}
\newcommand{\eea}{\end{eqnarray}}

\newcommand{\MS}{\overline{\rm MS}}
\begin{document}
\title
{Compelling evidence of renormalons in QCD from high order
perturbative expansions} 
\author{Clemens Bauer}
\affiliation{Institut f\"ur Theoretische Physik, Universit\"at
Regensburg, D-93040 Regensburg, Germany}
\author{Gunnar S.\ Bali}
\affiliation{Institut f\"ur Theoretische Physik, Universit\"at
Regensburg, D-93040 Regensburg, Germany}
\author{Antonio Pineda}
\affiliation{Grup de F\'{\i}sica Te\`orica, Universitat
Aut\`onoma de Barcelona, E-08193 Bellaterra, Barcelona, Spain}
\date{\today}
\begin{abstract}
We compute the
static self-energy of $\mathrm{SU(3)}$ gauge
theory in four spacetime dimensions to order $\alpha^{20}$
in the strong coupling constant $\alpha$.
We employ lattice regularization to enable a numerical simulation
within the framework of stochastic perturbation theory.
We find perfect agreement with the factorial growth of high order coefficients 
predicted by the conjectured renormalon picture based on
the operator product expansion.
\end{abstract}
\pacs{11.15.Bt,12.38.Cy,12.38.Bx,11.10.Jj,12.39.Hg}
\maketitle

Little is known about properties of quantum field theories
from first principles. This is
particularly so for asymptotically free gauge theories such as quantum
gluodynamics. One of the most salient features of this theory is the confinement
of charged objects. Yet this property has not been proven, and
the best evidence comes from the linearly rising 
static potential at large distances obtained  
in lattice simulations. Another expected property
is the asymptotic nature of perturbative weak coupling
expansions. In four dimensional non-Abelian
gauge theories one particular pattern of 
asymptotic divergence should be 
determined by the structure of the operator product
expansion (OPE). It is usually
named renormalon~\cite{Hooft} or, more specifically, infrared renormalon. 
Its existence has also not been proven but only tested 
assuming the dominance of $\beta_0$-terms, which amounts to
an effective Abelianization of the theory, 
or in the two dimensional $\mathrm{O}(N)$ model~\cite{David:1982qv}, 
where it is suppressed by 
powers of $1/N$. Moreover, the possible non-existence or irrelevance
of renormalons in Quantum 
Chromodynamics has been suggested in several papers, 
see, e.g.~\cite{Suslov:2005zi,Zakharov:2010tx} and references therein.
This has motivated dedicated 
high order perturbative expansions of the 
plaquette,
e.g.~\cite{DiRenzo:1995qc,Burgio:1997hc,Horsley:2001uy,Rakow:2005yn}, 
in lattice regularization, with conflicting conclusions.
Powers as high as $\alpha^{20}$
were achieved in the most recent simulation~\cite{Horsley:2010af}.
However, the expected asymptotic behaviour was not seen. 
If confirmed, this non-observation would 
cast doubt on the well-accepted lore of the
OPE and renormalon physics (see~\cite{Beneke:1998ui} for a comprehensive
review), and would significantly affect the phenomenological analysis
of data from high energy physics experiments on the decay of heavy
hadrons, heavy quark masses, the running coupling parameter, parton
distributions, etc.. Therefore, this issue should be clarified unambiguously.

In this letter we present compelling numerical evidence that the expected 
renormalons indeed exist not only in models but in real
gluodynamics.
We also argue why previous analyses based on the plaquette
have failed to detect them.
The vital and new ingredients of our study are as follows.\\
(a)
We consider a perturbative series whose leading renormalon is dictated
by a dimension $d=1$ operator, rather than by the $d=4$ plaquette.\\
(b) Using a higher order integrator and employing twisted boundary
conditions, among other improvements,
we are able to obtain results of unprecedented precision
on an extensive set of spacetime volumes.\\
(c) We carefully extrapolate to the infinite volume limit, thoroughly investigating finite size effects.

Perturbative expansions in powers of $\alpha$, 
\begin{equation}
K=\sum_n k_n\alpha^n,
\end{equation}
are believed
to be asymptotic and not Borel
summable in QCD, due to the
existence of singularities in the Borel plane (renormalons).
Typically $k_n$ will diverge like
$a_d^nn!$, with a constant $a_d$.
This divergence pattern of $k_n$ 
should not be arbitrary but consistent with the OPE
associated to a physical observable. Even though this factorial growth 
was originally discovered analyzing the Feynman 
diagrams that contribute to the large $\beta_0$ approximation, 
the correct divergent structure can only be inferred by assuming that
the perturbative series is asymptotic and complies with the OPE.
The OPE fixes the positions and the structure of the renormalon
singularities in the complex Borel plane, resulting in a
more intricate
pattern that cannot be obtained from the large $\beta_0$ approximation
alone.
Successive contributions~$k_n\alpha^n$ decrease for small
orders $n$ down to a minimum at $n_0\sim1/(|a_d|\alpha)$.
Higher order contributions
should be neglected and introduce an ambiguity of the order
of this minimum term,
$k_{n_0}\alpha^{n_0}\sim\exp[-1/(|a_d|\alpha)]$.

Within the OPE an observable $R(q,\Lambda)$
can be factorized into short distance
Wilson coefficients $C_i(q,\mu)$ and non-perturbative
matrix elements $\langle O_i(\mu,\Lambda)\rangle$ of dimension $i$:
\begin{equation}
R=C_0(q,\mu)\langle O_0(\mu,\Lambda)\rangle
+C_d(q,\mu)\langle O_d(\mu,\Lambda)\rangle\!\!\left(\frac{\Lambda}{q}\right)^d\!\!+\cdots\,.
\end{equation}
$q$, $\Lambda$ and $\mu$ denote a perturbative, low momentum and factorization scale, respectively,
so that $q\gg\mu\gg\Lambda$.
For the plaquette, $\langle O_0\rangle=1$ and the next higher non-vanishing operator is the dimension
$d=4$ gluon condensate. In this case, the perturbative expansion of $C_0$ cannot be more accurate
than $\mathcal{O}(\Lambda^4/q^4)$ which is exactly of the size of the
$k_{n_0}\alpha^{n_0}$ term since 
\begin{equation}
\label{eq:renormal}
\left(\frac{\Lambda}{q}\right)^d\simeq\exp\left(-\frac{1}{|a_d|\alpha}\right)\,,\quad\mbox{where}\quad a_d=\frac{\beta_0}{2\pi d}\,,
\end{equation}
with $\beta_0=11$.
The so-called
leading infrared renormalon
of this expansion cancels the ultraviolet ambiguity of the next order non-perturbative matrix element
so that the physical observable $R$ is well-defined.

From this discussion it is evident that we should
study series expansions with the smallest possible
$n_0$ or, equivalently, $d$.
For $d=1$  the perturbative expansion should start to diverge at 
an order $n_0$ that amounts to about one fourth of
that for the plaquette. 
This applies to the pole mass (see~\cite{Bigi:1994em,Beneke:1994sw}) and to the associated self-energy
of a static source, which we consider here.
The latter does not have a continuum limit, as it linearly depends on the ultraviolet regulator. Here we consider lattice regularization
with the Wilson gauge
action~\cite{Wilson:1974sk} and write the self-energy in the 
following way:
\begin{equation}
\delta m=\frac{1}{a}\sum_{n\geq 0}c_n\alpha^{n+1}(1/a)\,.
\end{equation}
$a^{-1}$, the inverse lattice spacing, provides the ultraviolet cut-off.
The large $n$ behaviour of the coefficients $c_n$ is regulator independent,
universal and equal to the asymptotic behaviour of the pole mass
up to $\mathcal{O}(e^{-1/n})$ terms (due to subleading renormalons):
\begin{align}
\label{generalm}
c_n \stackrel{n\rightarrow\infty}{=}&
N_{m}\,\left(\frac{\beta_0}{2\pi}\right)^n
\,\frac{\Gamma(n+1+b)}{
\Gamma(1+b)}
\\\nonumber
&
\times
\left(
1+\frac{b}{(n+b)}s_1+ \cdots
\right)\,.
\end{align}
The coefficients $b$ and $s_1$ were computed in \cite{Beneke:1994rs}. They read 
(see \cite{Pineda:2001zq} for details)
\be
b=\frac{\beta_1}{2\beta_0^2}\,,
\quad
s_1=\frac{1}{4\beta_0^3b}\left(\frac{\beta_1^2}{\beta_0}-\beta_2\right)
\,.
\ee
For a static source in the fundamental (triplet) representation the
normalization constant $N_m$ is exactly the same as for
the leading renormalon of a heavy quark pole mass. This renormalon
is also related to a renormalon of the singlet static potential
since these contributions cancel from
the energy $E(r)=2m+V(r)$~\cite{Pineda:1998id,Hoang:1998nz,Beneke:1998rk}.
For adjoint sources
it corresponds to a specific combination of pole mass and adjoint static
potential renormalons~\cite{Bali:2003jq}.
The factor $N_m$ is cancelled in the ratios
\be
\label{cnratio}
\frac{c_{n}}{c_{n-1}}\frac{1}{n} =\frac{\beta_0}{2\pi}
\left[1 +\frac{b}{n} - (1-b\,s_1)\frac{b\,s_1}{n^2}
+\mathcal{O}\left(\frac{1}{n^3}\right)
\right]
\,.
\ee

We obtain the expansion coefficients $c_n$
of the static energy  from the temporal Polyakov
line on hypercubic lattices. We investigate volumes of $N_T$ lattice points in the time direction
and spatial extents of $N_S$ points.
Formally we may introduce an anisotropy $a_t\neq a_s$. In this case
the lattice action, that is invariant under time or parity reversal,
agrees with the continuum action up to $\mathcal{O}(a_t^2,a_s^2)$ terms.
The temporal and spatial lattice extents in physical units are given by
$a_tN_T$ and $a_sN_S$, respectively, so that the
only dimensionless combinations consistent with the leading
order lattice artefacts are $a_t^2/(a_tN_T)^2=1/N_T^2$ and $1/N_S^2$.
Therefore, within perturbation theory, where we cannot dynamically generate
additional scales, the leading order lattice artefacts are
indistinguishable from $\mathcal{O}(1/N_T^2,1/N_S^2)$ finite size effects.

We choose periodic boundary
conditions in time and, to eliminate zero modes and to improve
the numerical stability, twisted boundary
conditions~\cite{tHooft79,Parisi83,Luscher86,Arroyo88} in all spatial
directions. The Polyakov line is defined by
\begin{equation}
L^{(R)}(N_S,N_T)=\frac{1}{N_S^3}\sum_{\mathbf n}\frac{1}{d_R}
\mathrm{tr}\left[\prod_{n_4=0}^{N_T-1}U^R_4(n)\right]\,,
\end{equation}
where $U^R_{\mu}(n)\approx e^{igA^R_{\mu}[(n+1/2)a]}
\in\mathrm{SU(3)}$ denotes a gauge link in representation
$R$, connecting the
sites $n$ and $n+\hat{\mu}$, $n_i\in\{0,\ldots,N_S-1\}$,
$n_4\in\{0,\ldots,N_T-1\}$
and $g=\sqrt{4\pi\alpha}$.
We implement triplet and octet representations $R$ of
dimensions $d_R=3$ and 8.
The link $U_4(n)$ appears within the
covariant derivative of the static action $\bar{\psi}D_4\psi$,
the discretization of which is not unique.
We use singly stout-smeared~\cite{Morningstar:2003gk}
(smearing parameter $\rho=1/6$)
covariant transporters instead of $U_4(n)$
as a second, alternative choice,
to demonstrate the universality of our findings.

\begin{table}
\caption{Lattice geometries. Volumes with boldface time
extents are expanded up to $\mathcal{O}(\alpha^{20})$, the
others up to $\mathcal{O}(\alpha^{12})$.\label{tab:volumes}}
\begin{center}
\begin{ruledtabular}
\begin{tabular}{cccc}
$N_S$&$N_T$&$N_S$&$N_T$\\\hline
7&\textbf{7}, \textbf{8}         &11&\textbf{16}\\
8&\textbf{8}, \textbf{10}, 12, 16&12&\textbf{12}\\
9&\textbf{12}                    &14&\textbf{14}\\
10&8, \textbf{10}, 12, 16, 20       &16&12, 16, 20 
\end{tabular}
\end{ruledtabular}
\end{center}
\end{table}

We remark that neither the lattice spacing nor the strong coupling
parameter $\alpha$ enter our simulations explicitly.
Numerical stochastic perturbation theory (NSPT)~\cite{DRMMOLatt94,DRMMO94,DR0}
enables us to directly calculate coefficients of
perturbative expansions.
We employ the variant of the Langevin algorithm introduced
in \cite{Torrero:08} that only quadratically depends on a time step
$\Delta\tau$.
Extrapolations to $\Delta\tau=0$ were performed on a subset
of lattice volumes where we found agreement within statistical
errors between all our extrapolated expansion coefficients
and those obtained at $\Delta\tau = 0.05$. For the geometries listed in
Table~\ref{tab:volumes} we restrict ourselves to this fixed value, which, 
within errors, effectively corresponds to $\Delta\tau=0$.

We expand the logarithm of the smeared and unsmeared
Polyakov lines in different representations to obtain the
corresponding static energies:
\begin{equation}
\label{eq:defP}
P(N_S,N_T)
=-\frac{\ln\langle L(N_S,N_T)\rangle}{aN_T}\
\stackrel{N_S,N_T\rightarrow\infty}{\longrightarrow} \delta m
\,.
\end{equation}
Fortunately, the dependence
of this logarithm on $N_T$ and $N_S$ can be deduced and
only a few parameters need to be fitted at each order:
\begin{align}\nonumber
aP&=\sum_{n\geq 0}
\left[
c_n\alpha^{n+1}\!\left(a^{-1}\right)
-\frac{f_n}{N_S}\alpha^{n+1}\!\!\left(\left(aN_S\right)^{-1}\right)
\right.
\\
\nonumber
&+\left.
{\cal O}\left(\frac{1}{N_T^2},\frac{1}{N_S^2}\right)
\right]\\
\nonumber
&\approx\sum_{n\geq 0}
\left[c_n+\Delta_n^{(1)}(N_S)+\Delta_n^{(2)}(N_S,N_T)\right]\!\alpha^{n+1}\!\!\left(a^{-1}\right)\,,
\\\label{eq:dlo}
\Delta_n^{(1)}&=-\frac{1}{N_S}\left[f_n+\mathrm{logs}^{(c)}_n(N_S)\right]\,,\\\nonumber
\Delta_n^{(2)}&=\frac{1}{N_T^2}\left\{v_n
-\frac{1}{N_S}\left[f_n^{(v)}+\mathrm{logs}^{(v)}_n(N_S)\right]\right\}\\\nonumber
&+\frac{1}{N_S^2}\left\{w_n
-\frac{1}{N_S}\left[f_n^{(w)}+\mathrm{logs}^{(w)}_n(N_S)\right]\right\}
\,.
\end{align}
The $\mathrm{logs}^{(c)}_n(N_S)$ are
polynomials of $\ln(N_S)$ of order $n-1$,
with coefficient functions
that depend on $f_j$ and the $\beta$-function
coefficients $\beta_j$ where $j\leq n-1$.
These terms are entirely determined
by the renormalization group running of
$\alpha$. 
The $\mathrm{logs}^{(v/w)}_n(N_S)$ are obtained in the same way.
In the $N_T\rightarrow\infty$ limit $\Delta_n^{(1)}$ is
the dominant correction while $\Delta_n^{(2)}$ 
includes the leading $\mathcal{O}(1/N_T^2,1/N_S^2)$
lattice artifacts discussed above.

The term $\Delta_n^{(1)}$  originates from interactions with mirror 
images, see also~\cite{Trottier:2001vj}. This effectively produces
a static potential between charges 
separated at distances $aN_S$, but without self-energies.
Therefore, we expect the high order behaviour
of $f_n$ and $c_n$ to be dominated by one and the same renormalon.
This can also be illustrated 
considering the leading dressed
gluon propagator $D(k)\propto 1/k^2$, where $k_4=0$.
With the (formal) ultraviolet
cut-off $1/a$ and an infrared cut-off $1/(aN_S)$
this can be written as (ignoring lattice corrections),
\begin{align}
\label{cutoffeq}
P&\propto
\int_{1/(aN_S)}^{1/a}\!\!\!\!dk\, k^2 D(k)\\
&\sim\frac{1}{a}\sum_nc_n\alpha^{n+1}\!\left(a^{-1}\right)-\frac{1}{aN_S}
\sum_nc_n\alpha^{n+1}\!\left((aN_S)^{-1}\right)\,,\nonumber
\end{align}
after perturbatively expanding $D(k)$. 
When re-expressing $\alpha((aN_S)^{-1})$ in terms of
$\alpha(a^{-1})$ we may consider two situations: \\
(a) $N_S>e^n$. In this limit the last 
term of Eq. (\ref{cutoffeq}) is exponentially suppressed in $n$
and the renormalon can directly be obtained
from a large order expansion of $aP$.\\
(b) $N_S<e^n$. The last term of Eq. (\ref{cutoffeq}) is
important and the renormalon cancels
order-by-order in $n$. 

In present-day numerical
simulations $N_S<e^n$, and the term $\Delta_n^{(1)}$ needs to be
taken into account, in combination with $c_n$. A similar
phenomenon was numerically observed for
the static singlet energy $E(r)=2m+V(r)$~\cite{Pineda:2002se,Bali:2003jq}.
This teaches us that to correctly identify
the renormalon structure of $\delta m$, it is compulsory to 
incorporate the $1/N_S$ corrections.
So far, in studies of high order perturbative expansions of the plaquette
the corresponding finite size terms have been neglected.
Our fits indeed yield $f_n\simeq c_n$ for large $n$,
in clear support of the renormalon dominance picture.

In the lattice scheme $\beta_0$, $\beta_1$ and $\beta_2$ are
known~\cite{Bode:2001uz}. The effects of higher $\beta_j$ start 
at $\mathcal{O}(\alpha^5)$, but this uncertainty in our parametrization quickly becomes negligible at high orders
where the coefficients $f_j$, governed by the $d=1$ renormalon,
will dominate. This can be quantified systematically in a
large $n$ analysis~\cite{BBP},
where any possible renormalon of the lattice $\beta$-function is
subleading ($d>1$). To check this assumption and to justify
the truncation at $\beta_2$
we have performed fits including $\beta_j$
for $j\leq 0, 1$ and 2 (see below).

Starting at $\mathcal{O}(\alpha^4)$, one may expect
additional finite size terms
$\propto \ln(N_T/N_S)/N_S$ from a possible
mixing of the antitriplet interaction between
mirror charges with sextet and higher representations, mediated
by ultrasoft gluons, in analogy
to the mixing of singlet and octet static potentials in
potential nonrelativistic QCD (pNRQCD)~\cite{Brambilla:1999qa}.
These terms are subleading from the
renormalon point of view ($d=3$).
Moreover, $aN_S$ provides an infrared cut-off to gluon
momenta so that one would only expect such contributions in the
limit $N_S\gg N_T$ that we do not investigate and,
indeed, we see no numerical evidence of them.

Our data are sensitive to the $1/N_T^2$ correction
terms within $\Delta_n^{(2)}$.
However, including $w_n$ or $f_n^{(w)}$ as additional fit parameters
did not significantly improve the $\chi^2$-value and so
we decided to omit the $1/N_S^2$ and $1/N_S^3$ terms.
Note that these contributions, if present, can numerically
easily be distinguished
from $1/N_S$ and become irrelevant at relatively small $N_S$, unlike
$\mathrm{logs}(N_S)/N_S$ terms.
\begin{figure}
\includegraphics[width=.45\textwidth,clip=]{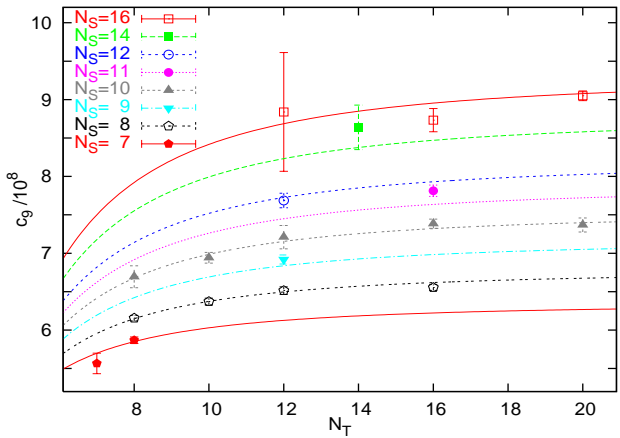}
\caption{Comparison between the global fit and data for $n=9$.
\label{fig:fit}}
\end{figure}
\begin{figure}[t]
\includegraphics[width=.45\textwidth,clip=]{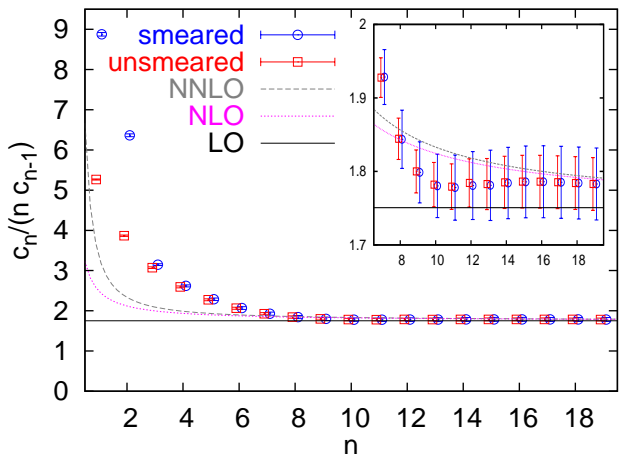}
\caption{The ratio
$c_n/(nc_{n-1})$ for the smeared and unsmeared fundamental static
self-energies, compared to the prediction
Eq.~(\protect\ref{cnratio}) at different orders of the
$1/n$ expansion.
\label{n20}}
\end{figure}
\begin{figure}
\includegraphics[width=.45\textwidth,clip=]{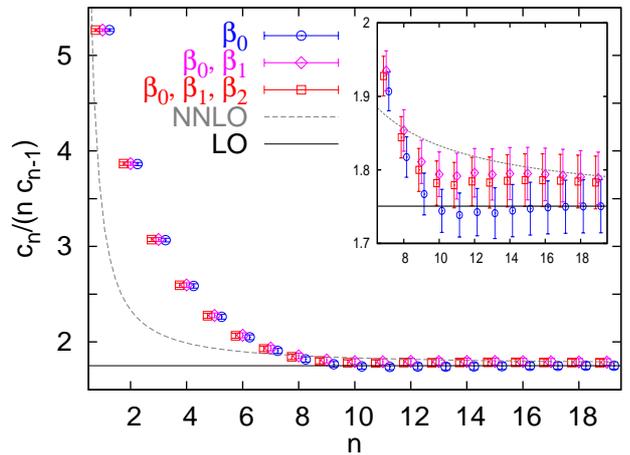}
\caption{The same as Fig.~\protect\ref{n20} for the unsmeared
data, truncating at different orders in $\beta_j$.
\label{betai}}
\end{figure}

As a cross-check
we calculate diagrammatically,
\begin{align}
c_0&=2.1172743570834807985970\ldots\,,\\
c_1&=11.1425(25)\,,\quad f_0=0.76962563284(2) \,,\\
\quad f_0^{(w)}&=0.14932(3)\,,\quad w_0=v_0=f_0^{(v)}=0\,,
\end{align}
for the unsmeared Polyakov line. In this case $f_0^{(w)}$, the
$1/N_S^3$ coefficient, does not vanish but it is small.
For fundamental sources, $c_0$ and $c_1$ were
known diagrammatically before and $c_2$
numerically~\cite{DiRenzo:2000nd,Trottier:2001vj}.
Our fit reproduces these values.
For adjoint sources the above coefficients need
to be multiplied by the factor $C_A/C_F=9/4$.

We exemplify the result of our global fit to the unsmeared
triplet data obtained on all our geometries (see Table~\ref{tab:volumes})
and orders of perturbation theory (with four parameters per order)
in Fig.~\ref{fig:fit}, where
a comparison to the $n=9$ data is shown.
We find smeared and unsmeared data to be well described by the
fits, with reasonable $\chi^2/N_{\mathrm{DF}}\approx 1.29$ and 1.46,
respectively. Note that the factorial growth found (and expected)
for the coefficients $f_n$ produces 
very sizable $1/N_S$ terms at high orders.

In Fig.~\ref{n20} we compare the infinite
volume extrapolated ratios $c_n/(nc_{n-1})$
to the theoretical prediction, Eq.~(\ref{cnratio}).
LO, NLO and NNLO refer to this prediction, truncated at $\mathcal{O}(1)$,
$\mathcal{O}(1/n)$ and $\mathcal{O}(1/n^2)$, respectively.
The data are robust to subtracting lattice artefacts
(the $\Delta^{(2)}$ terms of Eq.~(\ref{eq:dlo})) 
or to truncating at different orders in 
$\beta_j$, see Fig.~\ref{betai}. Particularly reassuring 
is the universality of the result; fits to smeared and unsmeared
Polyakov loop expansions
give the same large $n$ behaviour, fully consistent with the 
dominance and universality of the infrared renormalon;
smearing only affects the ultraviolet behaviour.
Fits to the octet representation data also show exactly the
same behaviour, again in agreement with the renormalon dominance picture. 
Also note that NSPT data for different orders are statistically correlated.
These correlations work in our favour.
We postpone the details of this to~\cite{BBP}.

Finally we determine the normalization of the pole mass renormalon,
see Eq.~(\ref{generalm}), and obtain $N^{\mathrm{lat}}_m=18.6(4)$
for the smeared and
$N^{\mathrm{lat}}_m=19.0(3)$ for the unsmeared static action.
Converting this to the modified minimal subtraction ($\MS$) scheme, we find
$N^{\MS}_m=\Lambda_{\mathrm{lat}}N^{\mathrm{lat}}_m/\Lambda_{\MS}=0.65(2)$.
This agrees remarkably well with the
estimate $N^{\MS}_m\simeq 0.62$ of~\cite{Pineda:2002se,Lee:2002sn} from
a $\MS$ scheme expansion 
up to ${\cal O}(\alpha^3)$,
in support of the claim that renormalon dominance starts
at much lower orders in the $\MS$ scheme (see for
instance \cite{Pineda:2001zq,Pineda:2002se,Bali:2003jq}). 
Preliminary results from directly converting our lattice data
to the $\MS$ scheme further reinforce this claim.
Irrespective of the scheme, the heavy quark pole mass can only
be defined up to 
an ambiguity of $\sim 0.65\,\Lambda_{\MS}$.
A more detailed analysis is in preparation~\cite{BBP}.

In conclusion, we have obtained the static self-energy
of $\mathrm{SU(3)}$ gauge theory in four spacetime dimensions
to $\mathcal{O}(\alpha^{20})$ in the lattice scheme.
For orders $n\gtrsim 9$ we find perfect agreement
with the factorial growth of the
coefficients, as predicted by the conjectured renormalon picture based on
the operator product expansion. Note that this implies that, in the
lattice scheme, we expect the 
renormalon dominance of the plaquette to set in at
values $n \gtrsim 4\times 9$ that so far have not been
realized in the literature.

\begin{acknowledgments}
We thank 
V.\ Braun, F.\ Di Renzo, M.\ Garc\'{\i}a P\'erez,
H.\ Perlt, A.\ Schiller and C.\ Torrero for discussions.
Computations were performed on Regensburg's Athene HPC cluster
and at the Leibniz Supercomputing Centre in Munich.~C.B.\ was supported by the Studienstiftung des deutschen
Volkes and by the Daimler und Benz Stiftung.
This work was supported by DAAD (Acciones Integradas
Hispano-Alemanas D/07/13355),
DFG SFB/TR 55, the
EU ITN STRONGnet grant 238353, the Spanish 
grants FPA2010-16963 and FPA2011-25948, and the Catalan grant SGR2009-00894.
\end{acknowledgments}

\end{document}